\documentclass[
  journal=pasa,
  manuscript=research-paper,
  year=2024,
  volume=XX,
]{cup-journal}

\usepackage{amsmath}
\usepackage[nopatch]{microtype}
\usepackage{booktabs}

\usepackage{siunitx}
\usepackage{amssymb}
\usepackage{multirow}
\usepackage{subcaption}
\usepackage{hyperref}
\usepackage{float}

\hypersetup{colorlinks=true,citecolor=blue,linkcolor=blue,urlcolor=blue}

\newcommand{\degree}[1]{${#1}^\circ$}

\title[Multi-band optical variability of blazar 1E 1458.8+2249]{Multi-band optical variability on diverse timescales of blazar 1E 1458.8+2249 
}

\author{Aykut Özdönmez}
\affiliation{Atatürk University, Faculty of Science, Department of Astronomy and Space Science, Yakutiye, 25240, Erzurum, Türkiye}
\email[Aykut Özdönmez]{aykut.ozdonmez@atauni.edu.tr}

\author{Murat Tekkeşinoğlu}
\affiliation{Atatürk University, Graduate School of Natural and Applied Sciences, Department of Astronomy and Astrophysics, Yakutiye, 25240, Erzurum, Türkiye}

\keywords{galaxies: general; galaxies: active; BL Lacertae objects: general; BL Lacertae objects: individual: 1E 1458.8+2249} 

\begin{document}

\begin{abstract}
This study presents an analysis of the optical variability of the blazar 1E 1458.8+2249 on diverse timescales using multi-band observations, including observations in the optical BVRI bands carried out with the T60 and T100 telescopes from 2020 to 2023 and ZTF gri data from 2018 to 2023. 
On seven nights, we searched for intraday variability using the power-enhanced F-test and the nested ANOVA test, but no significant variability was found.
The long-term light curve shows a variability behaviour in the optical BVRI bands with amplitudes of $\sim100\%$ and in the gri bands with amplitudes of $\sim120\%$, including short-term variability of up to $\sim1.1$ mag. 
Correlation analysis revealed a strong correlation between the optical multi-band emissions without any time lag.
From 62 nightly SEDs, we obtained spectral indices between 0.826 and 1.360, with an average of $1.128\pm0.063$.
The relationships of both spectral indices and colour with respect to brightness indicate a mild BWB trend throughout the observation period, both intraday and long-term.
We also performed a periodicity search using the Weighted Wavelet Z-transform (WWZ) and Lomb-Scargle methods. A recurrent optical emission pattern with a quasi-periodicity of $\sim340$ days is detected in the combined V- and R-band light curves. 
The observational results indicate that the blazar 1E 1458.8+2249 has a complex variability, while emphasising the need for future observations to unravel its underlying mechanisms.
\end{abstract}

\section{Introduction}
A subclass of Active Galactic Nuclei (AGNs), Blazars contain a relativistic jet closely aligned with the line of sight to the observer and a supermassive black hole (SMBH) with a mass of $10^{6-10}$ M\textsubscript{\(\odot\)} at the centre of the galaxy. They are characterised by rapid emission variations across wavelengths ranging from radio to gamma-ray,  high and variable polarisation, apparent superluminal motion, non-thermal continuous emission, and high-energy gamma-ray radiation \citep[e.g.,][]{1995PASP..107..803U, 2002ApJ...579..530W}. 

The multiwavelength spectral energy distribution (SED) of a blazar has a typical double hump structure \citep[e.g.,][]{1998MNRAS.299..433F}. The low energy bump peak ranges from IR/optical to UV/X-rays and is well explained by the synchrotron mechanism of electrons travelling at relativistic speed entering the magnetic field of the jet. The inverse Compton process satisfactorily reproduces the high-energy peak in the MeV-TeV range \citep[e.g.,][]{2010ApJ...716...30A, 2009ApJ...704...38S}. Blazars can also be divided into two subclasses: BL Lacertae objects (BL Lacs) and flat-spectrum radio quasars (FSRQs). FSRQs have broad emission lines in their optical spectra, whereas BL Lacs have very weak or no emission lines in their featureless optical spectra \citep{1996MNRAS.281..425M, 2012agn..book.....B}.

The nature of blazars can be studied through their variability, which is generally divided into intraday variability (IDV), short-term variability (STV), and long-term variability (LTV). The IDV, also known as microvariability, refers to flux variations over timescales ranging from tenths of minutes to hours, while the STV ranges from days to months, and the LTV has a longer timescale from months to years \citep[e.g.,][]{2004MNRAS.348..831X, 2008AJ....135.1384G, 2022ApJ...933...42A}. Various models related to jet and accretion disc have been proposed to explain the variability on diverse timescales, e.g., shocks moving within the jet \citep{1996A&AS..120C.537M}, gravitational microlensing \citep{1987A&A...171...49S}, and variations in the Doppler factor due to the spiral motion of the emission plasma \citep{2017Natur.552..374R}. However, many details of the models are still under discussion \citep{2021ApJ...923....7B}.
 
The colour and spectral changes associated with the variation of optical flux in blazars provide valuable information about their emission mechanisms. The bluer-when-brighter (BWB) trend is often observed in BL Lac objects. This trend is often explained by the shock-in-jet model and variations in the Doppler factor \citep[e.g.,][]{1997A&A...327...61G, 2002A&A...390..407V, 2010MNRAS.404.1992R, 2019MNRAS.488.4093A}. 
The redder-when-brighter (RWB) trend is more common in FSRQs and is thought to be caused by additional radiation from the accretion disc \citep[e.g.,][]{2006A&A...453..817V, 2011A&A...528A..95G}. Another colour trend is achromatic behaviour, which is often attributed to variations in the Doppler factor in the context of the geometrical scenario \citep{2002A&A...390..407V}. However, there are indications of more complex trends at different timescales \citep[e.g.,][]{2016MNRAS.458.1127G, 2017ApJ...844..107I, 2021A&A...645A.137A, 2022MNRAS.510.1791N}, and a consistent framework has not yet been established for blazars.

Across the electromagnetic spectrum, from radio to gamma rays, periodic or quasi-periodic oscillations (QPOs) of flux variations have been reported in some blazars \citep[e.g.,][]{ 1988ApJ...325..628S, 2001ApJS..134..181J,  2001A&A...377..396R, 2003A&A...402..151R, 2009ApJ...690..216G, 2009A&A...501..455V, 2014MNRAS.443...58W, 2017A&A...600A.132S, 2022MNRAS.513.5238R, 2023MNRAS.520.4118C, 2023ApJ...943...53K, 2024MNRAS.527.9132T}. Reported timescales vary from tens of minutes to days, weeks, and even decades. Several mechanisms have been proposed to explain year-long periodicities. These include the presence of a binary SMBH system, a plasma blob moving helically within the jet, and jet precession. Although the nature of the day-like QPOs is still uncertain, it is thought that they originate in the innermost central emission region, i.e., the accretion disc and the black hole.

1E 1458.8+2249 is a relatively bright BL Lac object with z = 0.235 \citep{1998A&A...329..853H}. However, there are only a few studies of this object including flux variation and SED since the first epoch \citep{1978ApJS...38..357F, 1990ApJ...348..141S}. The study by \citet{2003A&A...399...33M} analysed optical and X-ray observations of the blazar from 1994 to 2001. The blazar had an R-brightness between 15.5 and 16.5 mag between 1994 and 1998. In January 2000, a flare with a maximum of 14.75 mag was detected in the optical R band. On 8 February 2001, during the second flare, the brightest magnitude observed so far was $R=14.62$ mag. The spectral energy distribution studies showed that the SED classification of 1E 1458.8+2249 is a high energy peaked BL Lac object \citep[HBL; e.g.,][]{2004A&A...419...25F, 2023ApJS..268...23F}.  This indicates that the emission of 1E 1458.8 + 2249 is of synchrotron origin. \citet{2022MNRAS.510.1791N} studied the colour variations of 897 blazars using multiband light curves obtained with the Zwicky Transient Facility (ZTF). They reported that 1E 1458.8 + 2249 exhibits BWB behaviour. This strong BWB trend ($r\sim0.8$) between $V-I$ colour and $V$ magnitude was also reported by \citet{2023MNRAS.520.4118C} in their intraday (ID) observations. The authors detected that one of the IDV light curves exhibited a drastic outburst with a variability amplitude of 0.97 mag ($\%96.5$) on 26 May 2011, while the other IDV on 14 April 2020 showed a variability amplitude of only 0.1 mag ($\%8.7$).

In this study, we present multi-colour observations of the blazar 1E 1458.8+2249 to investigate optical flux variations, correlations between different optical bands, and colour behaviour over intraday to long-term timescales. By combining ZTF and our observations, we analyse the quasi-periodic variability behaviour of the blazar. 

\section{Data}\label{sec:obs_data}
Optical observations were performed in the BVRI bands using a 1.0 m RC telescope (T100) and a 60 cm RC robotic telescope (T60) at the Tubitak National Observatory (TUG) from May 2020 to February 2023. The T100 telescope is equipped with a cryogenic cooled CCD (model SI 1100 Cryo, size 4096$\times$4037 px). The T60 is equipped with a thermally cooled CCD (model Andor iKon-L, size 2048$\times$2048 px). The exposure times were determined based on the photometric band and the brightness of the source, with a range of 20 to 180 seconds. A total of 1555 $BVRI$ frames were acquired over 169 nights, including seven nightly follow-up observations. Standard data reduction and calibration were performed on all CCD frames, i.e., bias subtraction, twilight flat-fielding, cosmic ray removal, and aperture photometry. In order to evaluate the instrumental magnitude of the blazar, we employed star A as the reference, as its position, brightness, and colour are closest to that of the blazar, and stars C1, C2, and C3 as comparisons\footnote{www.lsw.uni-heidelberg.de/projects/extragalactic/charts/1458.8+228}, as provided by \citet{1998PASP..110..105F}. It is important to note that only instrumental magnitudes were used for the B-band data since there are no B-band magnitudes for the standard stars. Therefore, the B-band light curves were not included in the analyses that require calibrated magnitudes.
We provide our observational data, including instrumental magnitudes, exposure times, and dates, in the online supplementary data.

In addition to our observational data, we collected light curves in the $gri$ bands from the ZTF database\footnote{www.ztf.caltech.edu/}, spanning from March 2018 (MJD 58194) and July 2023. In order to maintain the quality and goodness of the ZTF light curve, we selected 1,180 ZTF data points with catflags = 0 and chi < 4.

\section{Intraday, short-term, and long-term Variability}
\begin{figure*}[bt!]
\includegraphics[width=0.85\linewidth]{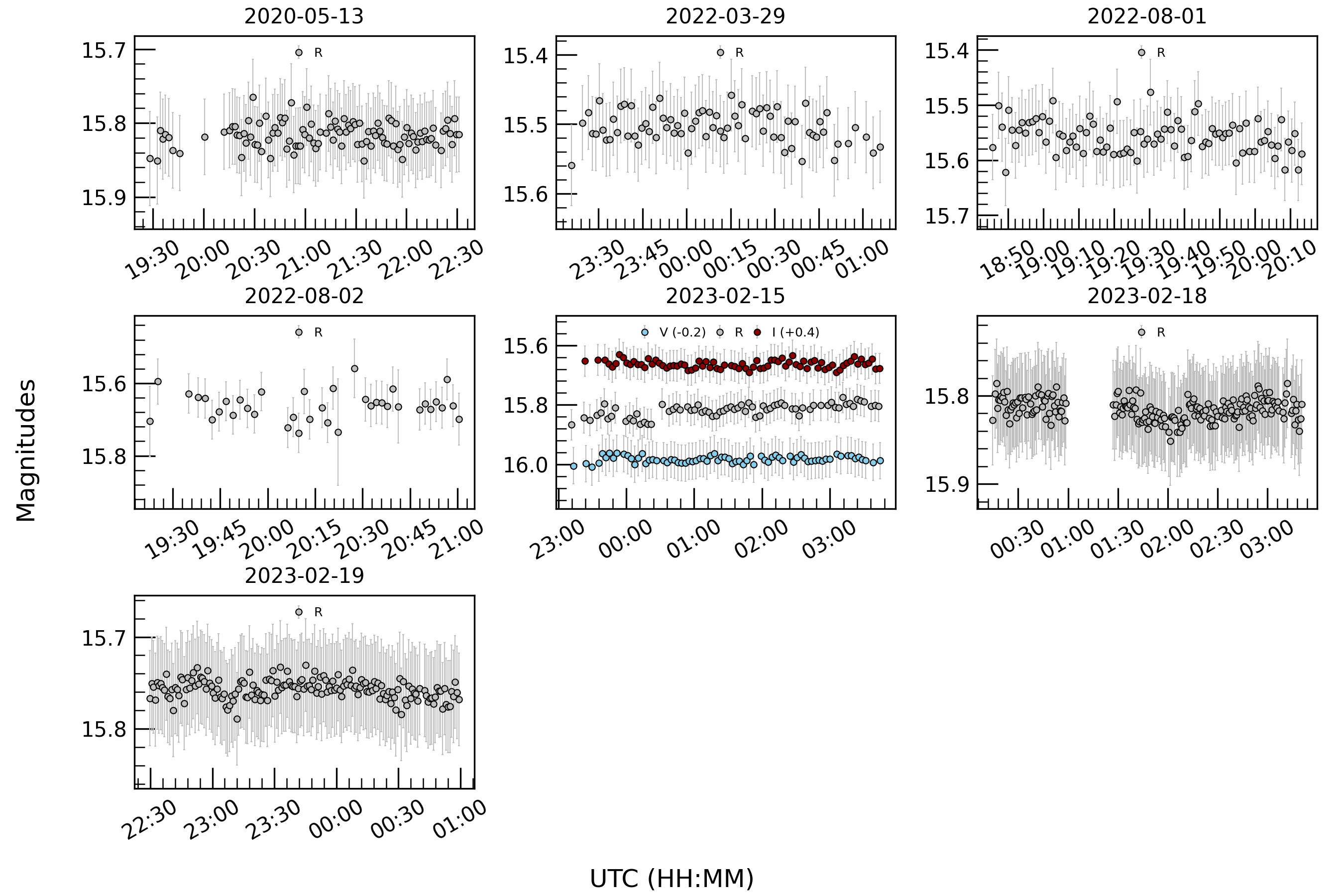}
\caption{Intraday light curves of the blazar 1E 1458.8+2249 in the optical $VRI$ bands. The blue, gray, and dark red circles represent the V-, R-, and I-band magnitudes, respectively, with a labelled offset. The dates of the observations are given at the top of each plot.}
\label{fig:IDV_lc}
\end{figure*}

To understand the characteristics of the emission regions and the underlying radiative processes, it is essential to study the flux variability of blazars at different frequencies and timescales, from the shortest to the longest. Our quasi-simultaneous optical observations provide the opportunity to study the variability of the blazar from intraday to long-term scales. In order to obtain a meaningful IDV, we have made seven nightly observations of at least one hour duration, i.e., six of them in the optical R-band and one of them in $BVRI$. Figure \ref{fig:IDV_lc} shows the light curves for these seven observation nights.

To search for variability in the intraday light curves of the blazar 1E 1458.8+2249, two widely used statistical tests were employed: the power-enhanced F-test and the nested ANOVA (analysis of variance) test \citep[e.g.,][]{2015MNRAS.452.4263G, 2019ApJ...871..192P, 2022ApJ...933...42A, 2020ApJ...900..137W}. As noted by \citet{2015AJ....150...44D}, other commonly used statistical tests to study optical variability, such as the C-test, F-test, and chi-square test, have several limitations that can lead to less reliable results. To improve the robustness of the analyses, it is crucial to include differential light curves from several comparison stars.

The power-enhanced F-test is defined as the comparison of the variance of the blazar's light curve with the combined variance of multiple comparison stars, as $F_{enh} = \frac{s_{bl}^2}{s_c^2}$ \citep{2014AJ....148...93D}. 
The differential variance of the blazar is represented by $s_{bl}^2$, while $s_c^2$ is the combined variance of the comparison stars.  The degrees of freedom (DOF) are formulated as $u_{bl} = N-1$ and $u_c = k(N-1)$. The critical value ($F_{critical}$) for the variability test is calculated at a 99\% confidence level ($\alpha = 0.01$). If $F_{enh}$ is greater than or equal to $F_{critical}$, the blazar variability hypothesis is valid. In our research, we used differential light curves of the reference star (the field star A) and three comparison stars to estimate $F_\text{enh}$ values and compare them with the critical value ($F_c$) for each observation day.

Nested ANOVA, which is a modified version of ANOVA, is used as a statistical tool to assess mean variations between groups in the differential light curves of the blazar generated by using multiple reference stars. An advantage of nested ANOVA is that it is independent of a specific comparison star, allowing all available stars to be used as reference stars in the analysis.
In our study, four standard stars (A, C1, C2, and C3) were used to construct differential LCs of the blazar. These differential LCs were divided into groups of five data points each. Following the methodology in \citet{montgomery2012design} and \citet{2015AJ....150...44D}, the mean square of the groups ($MS_G$) and the nested observations within the groups ($MS_{O_{G}}$) were calculated. The resulting ratio, $F = MS_G/MS_{O_{G}}$, follows an $F$ distribution with $(a-1)$ and $a(b-1)$ degrees of freedom in the numerator and denominator, respectively. The variability of the blazar's light curve was tested for a significance level of $\alpha = 0.01$.

The results of the $F_\text{enh}$-tests and the nested ANOVA tests are listed in Table \ref{tab:IDV_vartest}. We have assumed that the light curve is variable (V) if the $F$-statistic equals or exceeds the critical value ($F_c$) in both tests; otherwise, it is non-variable (NV). Although the daily monitoring time ranges from 1 to 5 hours, no variability on minute timescales was found in the seven intraday LCs.

\begin{table*}[bt!]
\caption{The test results of the IDV of the blazar S5 0716+714.} 
\label{tab:IDV_vartest}
\centering
\begin{tabular}{lcccccccccc}
\hline
\hline
Obs. date &  Band & Average & $t_\text{obs}$ & \multicolumn{3}{c}{Power-enhanced F-test}  &  \multicolumn{3}{c}{Nested ANOVA test}& Status \\
\cline{5-7}\cline{8-10}
 yyyy-mm-dd &   & Magnitude & hours &DOF($\nu_1$,$\nu_2$) &  $F_{enh}$ &  $F_c$ & DOF($\nu_1$,$\nu_2$) &  $F$ &  $F_c$  \\
\hline
2020-05-13 &  R &     15.82 &      3.05 &     (108, 324) &    0.85 &        1.42 &         (20, 84) &     0.64 &        2.10 &     NV \\
2022-03-29 &  R &     15.50 &      1.75 &      (74, 222) &    0.01 &        1.53 &         (14, 60) &     0.68 &        2.39 &     NV \\
2022-08-01 &  R &     15.56 &      1.46 &      (90, 270) &    0.72 &        1.47 &         (17, 72) &     1.72 &        2.23 &     NV \\
2022-08-02 &  R &     15.66 &      1.63 &      (37, 111) &    0.51 &        1.80 &          (6, 28) &     0.70 &        3.53 &     NV \\
2023-02-15 &  B &     -     &      4.14 &      (68, 204) &    0.66 &        1.55 &         (12, 52) &     1.24 &        2.55 &     NV \\
2023-02-15 &  V &     16.18 &      4.52 &      (73, 219) &    0.43 &        1.53 &         (13, 56) &     1.40 &        2.47 &     NV \\
2023-02-15 &  R &     15.82 &      4.52 &      (70, 210) &    0.96 &        1.54 &         (13, 56) &     5.46 &        2.47 &     NV \\
2023-02-15 &  I &     15.26 &      4.33 &      (77, 231) &    0.79 &        1.51 &         (14, 60) &     1.65 &        2.39 &     NV \\
2023-02-18 &  R &     15.81 &      3.10 &     (207, 621) &    0.83 &        1.29 &        (40, 164) &     2.05 &        1.72 &     NV \\
2023-02-19 &  R &     15.76 &      2.49 &     (169, 507) &    0.75 &        1.33 &        (33, 136) &     4.11 &        1.81 &     NV \\
\hline
\end{tabular}
\end{table*}

\begin{figure*}[bt!]
\includegraphics[width=0.8\linewidth]{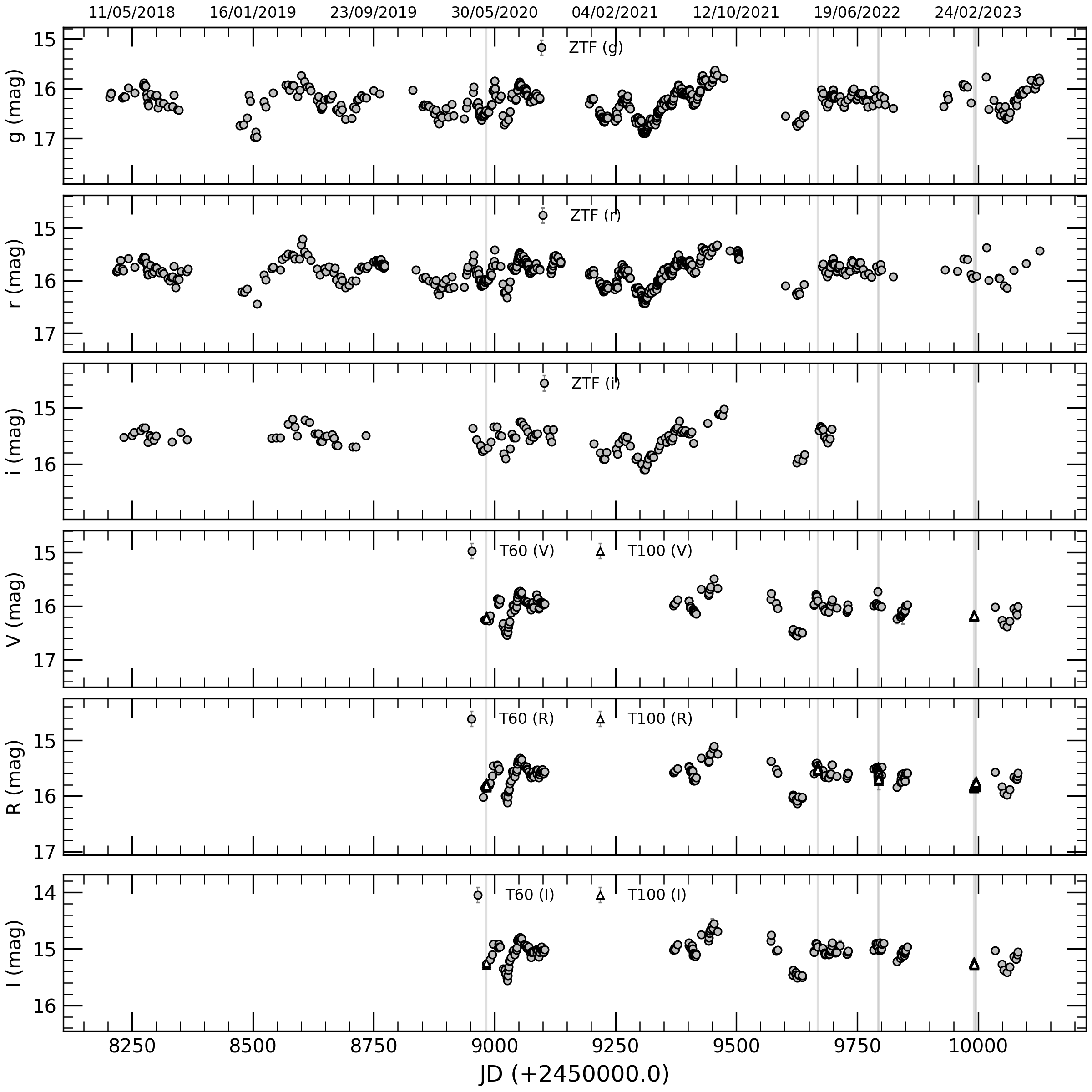}
\caption{Long-term light curves of the blazar 1E 1458.8+2249 in the optical gri bands and $VRI$ bands. Vertical lines represent the date of the intraday observation.}
\label{fig:LTV_BVRI}
\end{figure*}

Using all the observational data, we have constructed the long-term optical light curve for both the ZTF gri and VRI bands, as shown in Figure \ref{fig:LTV_BVRI}. 
Table \ref{tab:LTV_result} gives a brief overview of the LTV light curves, consisting of the minimum, maximum, and mean magnitudes. In addition, we have calculated the variability amplitude ($A$) as defined by \citet{1996A&A...305...42H}. It is defined as $A = \sqrt{(A_{\text{max}} - A_{\text{min}})^{2} - 2\sigma^{2}}$, where $A_{\text{max}}$ and $A_{\text{min}}$ are the maximum and minimum magnitudes of the LTV light curve, respectively, and $\sigma$ is the mean error.
The long-term light curve shows variability with amplitudes ranging from 100\% to 133\% depending on the band. During our observations, we recorded the brightest magnitude as R=15.109 mag on MJD 59453, while the blazar was at its faintest with R=16.136 mag on MJD 59625. The mean R-band magnitude during the monitoring period was $15.716\pm0.052$, and the variability amplitude was calculated as 102.397\%. 
ZTF observations from 2018 to 2023 show the brightest magnitude at $r = 15.320$ and the faintest magnitudes at $r = 16.427$, giving an average magnitude of $15.829$ and a variability amplitude of $122\%$. During these 5 years of observation, no flare was detected, and the brightness of the blazar didn't reach the maximum brightness ($R\sim 14.6$) which as in February 2001. 
The short-term light curve shows a continuous brightening from 2021 April to September (over $\sim150$ days) by $\sim1.1$ mag in the ZTF $gri$ bands. The LTV light curves, especially between MJD 8900 and 9300, show fluctuations on shorter timescales of the order of days with a smaller amplitude of up to 0.8 mag.

\begin{table*}[bt!]
\caption{Results of the LTV analysis of the blazar.}
\label{tab:LTV_result}                   
\centering 
\begin{tabular}{cccccc}       
\hline\hline                		 
Band &  Brightest magnitude/MJD & Faintest magnitude/MJD  & Average magnitude & Variability amplitude (\%)\\
\hline
g  & $15.634\pm0.014$ /  59455.69269 & $16.897\pm0.018$ / 59311.87397 & $16.282\pm0.016$ & 133.583\\
r  & $15.320\pm0.013$ /  59460.703131 & $16.427\pm0.016$ / 59311.96738 & $15.829\pm0.014$ & 122.785\\
i  & $15.112\pm0.013$ /  59462.66675 & $16.094\pm0.016$ / 59311.84264 & $15.552\pm0.014$ & 107.562\\
V  & $15.490\pm0.063$ /  59453.23145 & $16.548\pm0.064$ / 59625.61332 & $16.093\pm0.063$ & 105.431\\
R  & $15.109\pm0.051$ /  59453.23300 & $16.136\pm0.053$ / 59625.61202 & $15.716\pm0.052$ & 102.397\\
I  & $14.552\pm0.051$ /  59453.23401 & $15.560\pm0.053$ / 59026.28456 & $15.126\pm0.053$ & 100.572\\
\hline                          
\end{tabular}
\end{table*}

\section{The multiband colour/spectral behaviour and the correlation analysis}
\begin{figure}[bt!]
    \centering
    \includegraphics[width=0.9\linewidth]{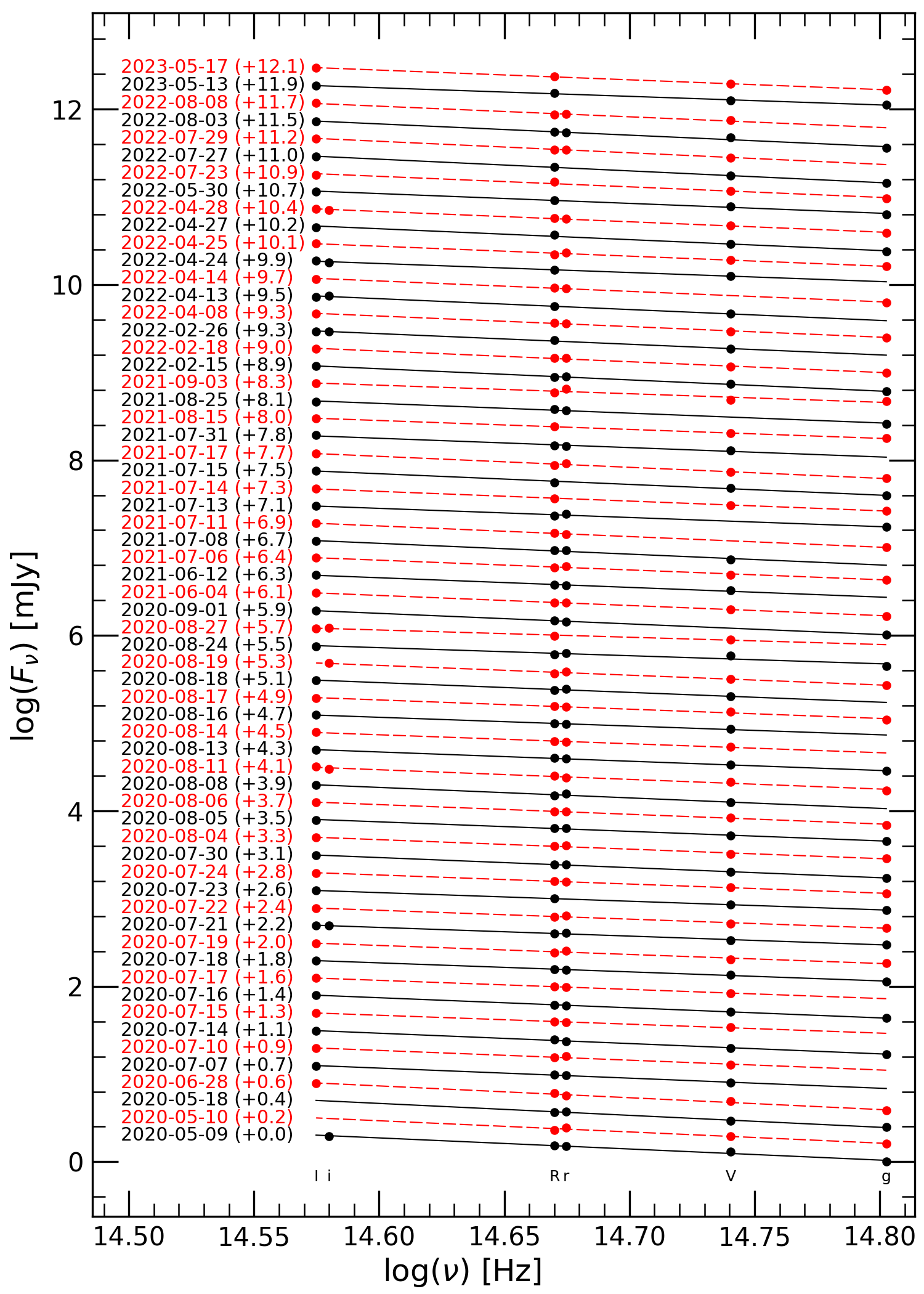}
    \caption{The intranight SEDs of the blazar 1E 1458.8+2249 in optical multi-bands.}
    \label{fig:SEDs}
\end{figure}

\begin{figure}[bt!]
    \centering
    \includegraphics[width=0.8\linewidth]{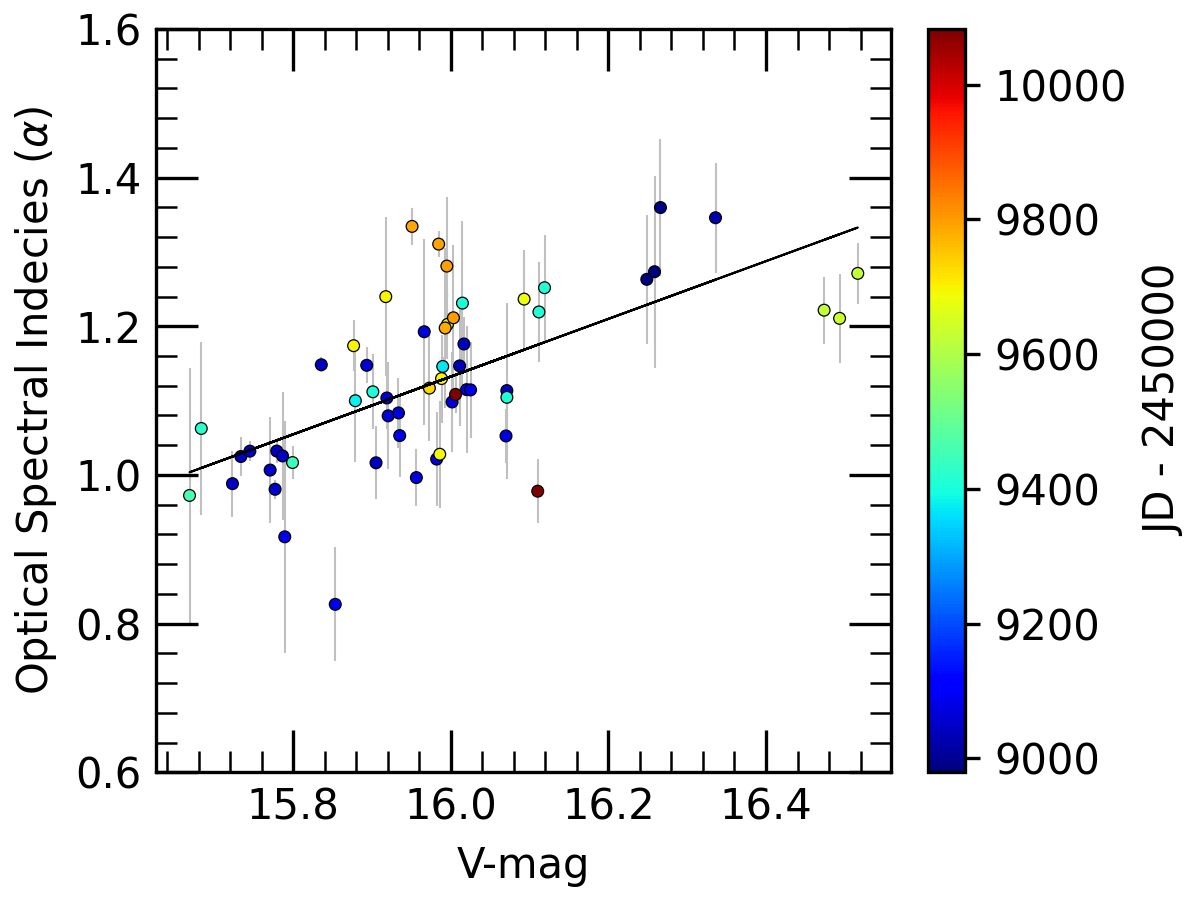}
    \caption{Variation of spectral index with V-mag during the entire monitoring period. The colour of the dots indicates the time of observation.}
    \label{fig:alpha_Vmag}
\end{figure}

Quasi-thermal emission from the accretion disc and non-thermal synchrotron emission from the relativistic jet drive the rapid flux and spectral variations in blazars. Understanding the spectral variations of blazars is crucial for identifying the different components that influence the observed flux. 
To examine the daily multiband optical spectral energy distributions (SEDs) of the blazar 1E 1458.8+2249, we used nearly simultaneous observations consisting of at least the four $VRI$- and $gri$-bands. The calibrated magnitudes of the blazar were corrected for the Galactic extinction using the $A_\lambda$ values taken from the NASA/IPAC Extragalactic Database\footnote{https://ned.ipac.caltech.edu}. Using the zero-points in \citet{1998A&A...333..231B} for the $VRI$-bands and in the filter profile service of the Spanish Virtual Observatory (SVO)\footnote{http://svo2.cab.inta-csic.es/theory/fps/index.php?id=Palomar/ZTF} for the ZTF $gri$-bands, these extinction-corrected magnitudes were then converted into corresponding fluxes. The optical spectral energy distributions (SEDs) were computed for the 62 nights, as shown in Figure \ref{fig:SEDs}. The SEDs were found to follow a simple power law relationship of the form $F_{\nu} \propto \nu^{-\alpha}$, where $\alpha$ represents the optical spectral index. The spectral indices for each night were determined by fitting a linear model of $log(F_{\nu}) = -\alpha~log(\nu)+ C$. The analysis reveals that the spectral indices range from 0.826 to 1.360, with an average of $1.128 \pm0.063$. The standard deviation is calculated to be 0.11. Figure \ref{fig:alpha_Vmag} displays the correlation between the spectral index and V-band brightness. Linear regression analysis shows that the correlation has a slope of $0.333\pm0.048$ (see Table \ref{tab:LTV_cmd}), i.e., a mild BWB trend, but there is no significant trend in $\alpha$ over time.

\begin{figure}[bt!]
    \centering
    \includegraphics[width=0.8\linewidth]{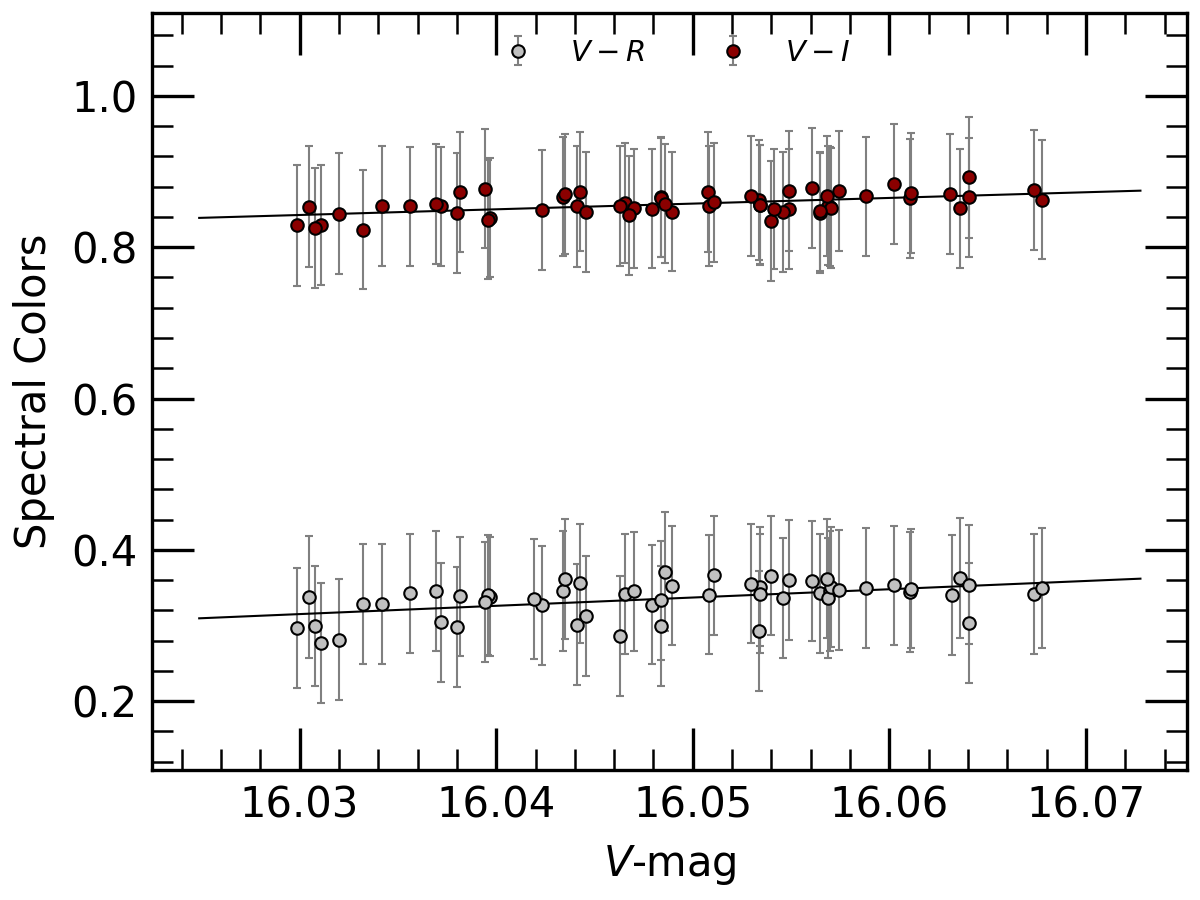}
    \caption{The intraday colour–magnitude plots of the blazar 1E 1458.8+2249 for the date 15 Feb 2023. The silver and dark red dots represent the $V-R$ and $V-I$ colour indices. Black lines represent linear fits. }
    \label{fig:IDV_cmd}
\end{figure}

Since the colour variability of the blazars reflects their spectral variability, we performed a correlation analysis between brightness and colours on both intraday and long-term timescales. We have plotted the $V-I$ and $V-R$ colour indices against the $V$ magnitude shown in Figure \ref{fig:IDV_cmd}, and the results of the linear regressions are listed in Table \ref{tab:IDV_cmd_fit}. We found mild correlations ($r\sim0.5$ and $p<0.05$) for both the $V-R$ and $V-I$ colour indices, indicating a non-significant BWB trend taking into account the large colour errors with an average of 0.079 mag.

\begin{table}
\centering
\caption{Correlation between the colour indices and V-band magnitude of multiband IDV data.}
\label{tab:IDV_cmd_fit}
\scriptsize
\begin{tabular}{lccccc}
\hline
Date &   colour &             m &             c &      r &         p \\
\hline
2023-02-15 &   $V-R$ & $1.09\pm0.27$ & $ -17.23\pm4.30$ & $+0.49$ & 1.47e-04 \\
 &   $V-I$ & $0.75\pm0.16$ & $ -11.21\pm2.52$ & $+0.53$ & 1.23e-05 \\
\hline
\end{tabular}
\end{table}

\begin{figure}[bt!]
\includegraphics[width=0.9\linewidth]{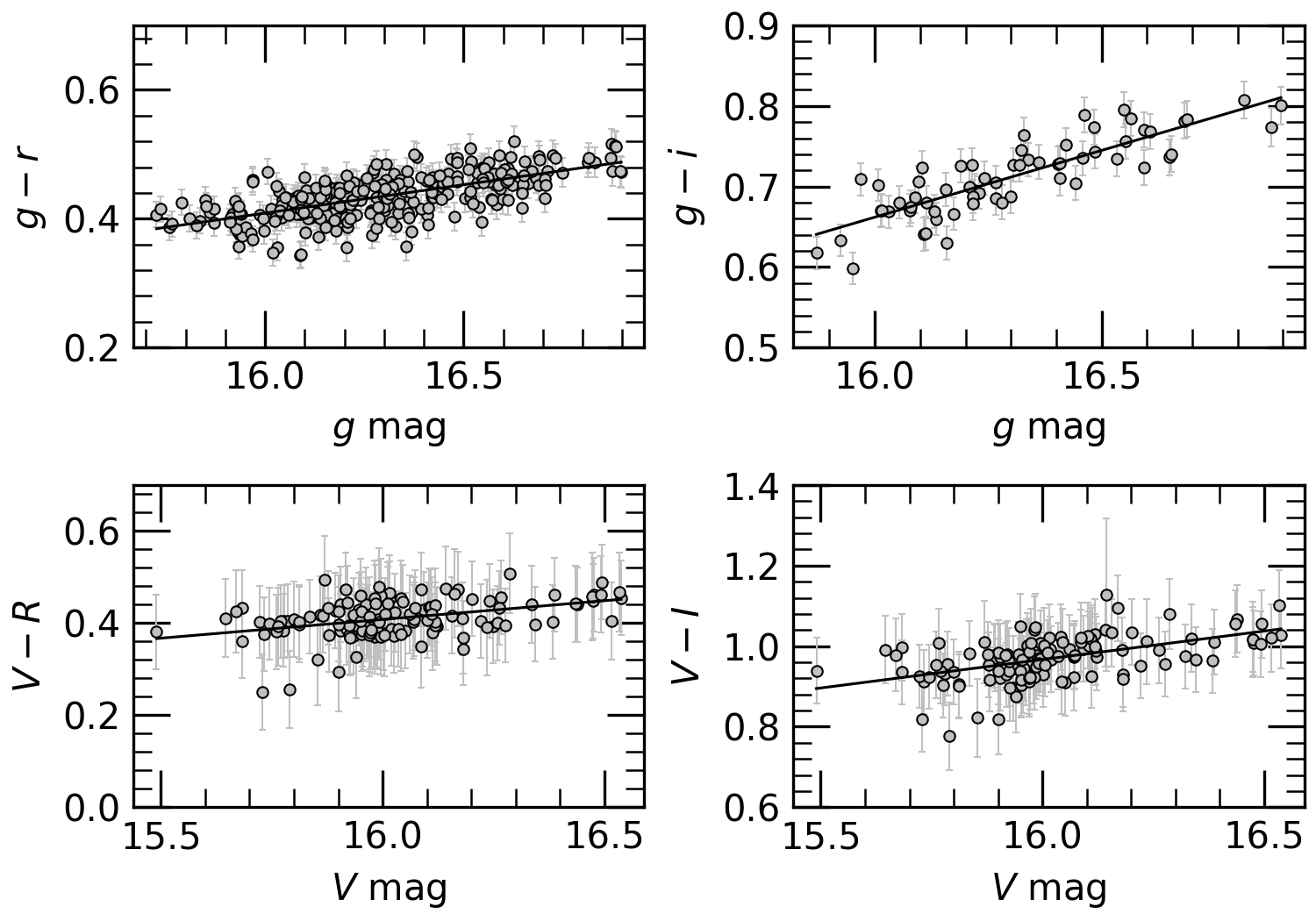}
\caption{colour-magnitude diagram of the blazar 1E 1458.8+2249 on long timescale. Black lines represent the linear fits.}
\label{fig:LTV_cmd}
\end{figure}

To investigate the long-term colour variations of the blazar, we analysed our $VRI$ and ZTF gri data sets separately.
We computed colour indices by coupling data from the same night. Therefore, we plot the colour indices with larger frequency bases, such as the $V - I$ and $V-R$ colours, against the $V$-band magnitude in Figure \ref{fig:LTV_cmd}. We have also plotted the $g-i$ and $g-r$ colours against the $g$-band magnitude for the ZTF dataset. 
We have listed the results of linear regression on the colour-magnitude relations in Table \ref{tab:LTV_cmd}. We found a strong correlation between ZTF $g-i$ colour and $g$ band magnitude with a correlation coefficient of $r=+0.84$, while moderate correlations ($r=0.66$) were found with the ZTF $g-r$ colour index. However, the colour relations of our data set have weaker correlations with $r\sim0.5$.

\begin{table}
\centering
\caption{The results of the linear regression between the spectral index ($\alpha$) / colours and the corresponding magnitude on the long timescales.}
\label{tab:LTV_cmd}
\scriptsize
\begin{tabular}{lcccc}
\hline
Model &  slope &  intercept &  r-value &   p-value \\
\hline
$V$ vs $\alpha$ & $+0.333\pm0.048$ & $-4.206\pm0.768$ & +0.535 & 2.134e-10 \\
$g$ vs $g-r$ & $+0.088\pm0.006$ & $-1.002\pm0.099$ & +0.655 & 9.563e-36 \\
$g$ vs $g-i$ & $+0.166\pm0.013$ & $-1.991\pm0.220$ & +0.844 & 3.595e-18 \\
$V$ vs $V-R$ & $+0.081\pm0.016$ & $-0.894\pm0.261$ & +0.403 & 1.785e-06 \\
$V$ vs $V-I$ & $+0.141\pm0.020$ & $-1.294\pm0.328$ & +0.527 & 2.272e-10 \\
\hline
\end{tabular}
\end{table}

Understanding the correlations between different spectral bands is crucial as any time lag identified implies spatial distinction between the corresponding emission regions. These results provide valuable insights into the physical processes taking place within the blazar, allowing a nuanced understanding of its complex dynamics and geometry.
The Discrete Correlation Function (DCF) is a powerful statistical tool for investigating unevenly distributed time series of multiband light curves \citep[][and references therein]{2015MNRAS.450..541A}.
The calculation of the DCF involves the use of the unbinned DCF (UDCF), expressed as:
\begin{equation}
UDCF_{ij}(\tau) = \frac{(a_i - \bar{a}) (b_j - \bar{b})}{\sqrt{(\sigma_{a^2} - e_{a^2})(\sigma_{b^2} - e_{b^2})}}
\end{equation}
Here, $\bar{a}$ and $\bar{b}$ are the mean values of the respective time-series datasets, while $\sigma_{a,b}$ and $e_{a,b}$ are their standard deviations and errors. The time delay between two data points is denoted by $\Delta t_{ij} = (t_{bj} - t_{ai})$. The DCF is then found by taking average of the UDCF values over the interval $\tau - \frac{\Delta\tau}{2} \leq \tau_{ij} \leq \tau + \frac{\Delta\tau}{2}$, using the methodology suggested by \citet{1988ApJ...333..646E}:
\begin{equation}
DCF(\tau) = \frac{\sum_{k=1}^{m} \text{UDCF}_k}{M}
\end{equation}
Here, "M" is the number of pairwise time lag values within the specified $\tau$ interval.

\begin{figure}[bt!]
\includegraphics[width=0.85\linewidth]{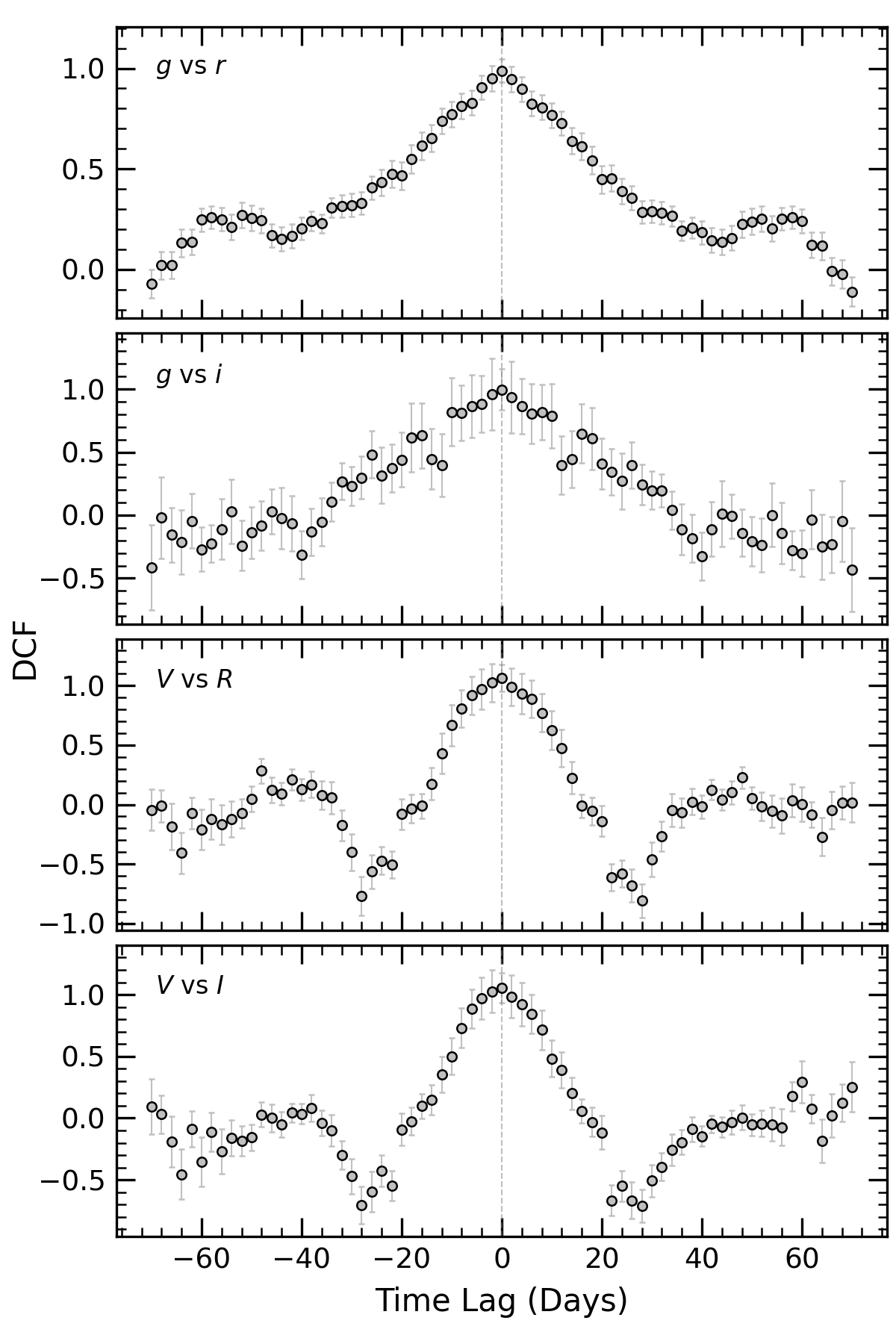}
\caption{Cross-correlation analysis of light curves in the ZTF $gri$ and optical $VRI$ bands using Discrete Correlation Function for the entire monitoring period.}
\label{fig:LTV_dcf}
\end{figure}

In the examination of the long-term light curves, weighted mean magnitudes and mean Julian Dates (JD) were computed from nightly binned observations. The DCF analyses (see Figure \ref{fig:LTV_dcf}) were applied to each pair combination of the nightly binned ZTF $gri$ and optical $VRI$ long-term light curves. Strong correlations were found between all multi-band pairs using a time binning value of 2 days, and no time lag was detected. 

\section{Optical Quasi-Periodic Oscillations}
\begin{figure}[bt!]
\includegraphics[width=0.9\linewidth]{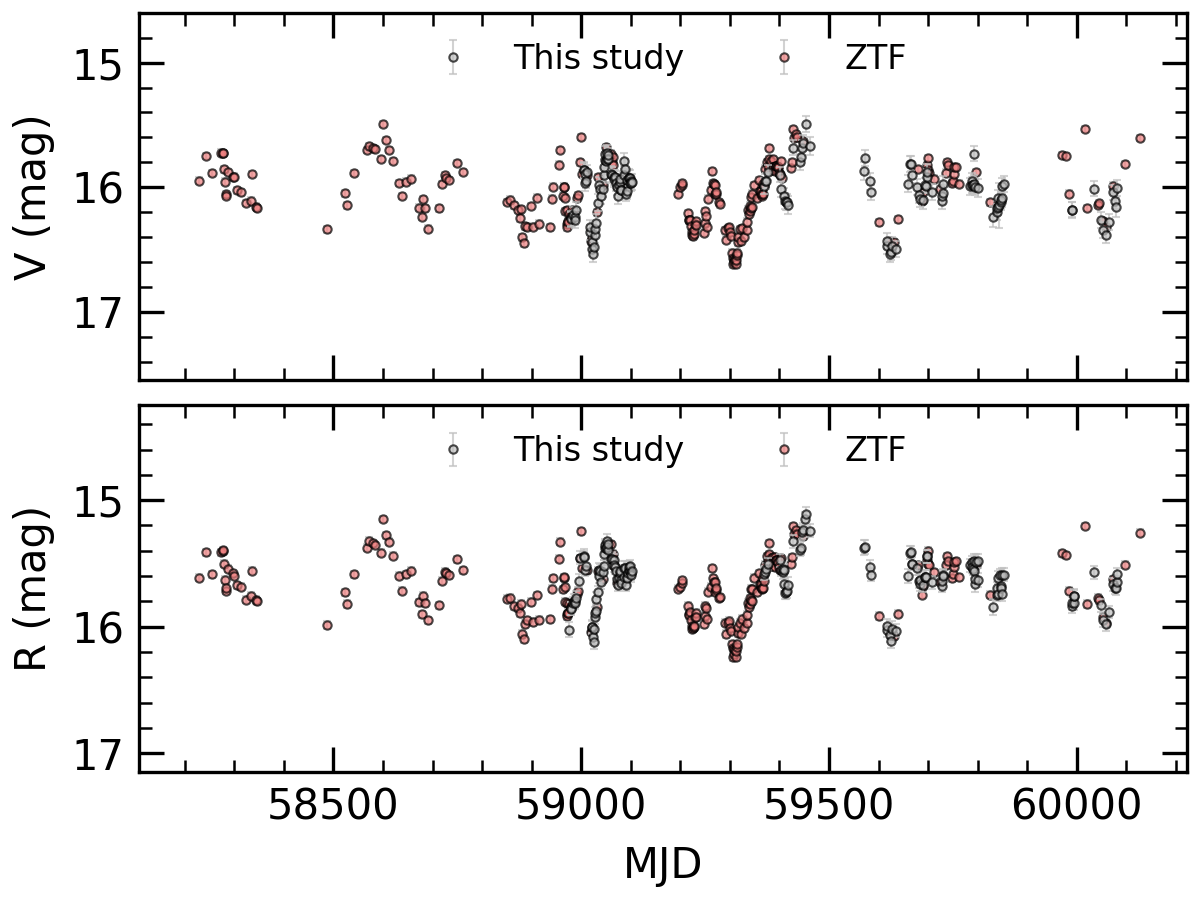}
\caption{The $V$- and $R$-band light curve combined from the calibrated ZTF data and our observations.}
\label{fig:LTV_combined}
\end{figure}

The availability of continuous observational data determines the accuracy of the quasi-periodic analysis. 
Thus, to fill gaps in the optical light curve and extend it, we transformed ZTF magnitudes to the V- and R-bands using the transformation equations\footnote{www.sdss4.org/dr17/algorithms/sdssUBVRITransform/} 
\citep{2005AAS...20713308L}. 
The resulting combined light curves, including both the transformed ZTF magnitudes and those obtained in our study, are shown in Figure \ref{fig:LTV_combined}. It is worth noting that \citet{2022MNRAS.516.3650Z} was able to identify quasi-periodicity in the quasar SDSS J1321+033055, even without taking into account the difference in magnitude between the CSS (Catalina Sky Survey) light curve and the ZTF light curves.

To identify any underlying periodicity in the combined light curve of 1E 1458.8+2249, we employed two widely used methods: the Lomb-Scargle (LS) periodogram \citep{1976Ap&SS..39..447L, 1982ApJ...263..835S}, and  
the Weighted Wavelet Z-transform \citep[WWZ;][]{1996AJ....112.1709F}.

The LS periodogram is a Fourier transform for detecting periodicity in unevenly distributed time series data. It uses likelihood minimisation to integrate sinusoidal components into the time series data.
Utilising the Python Astropy package\footnote{https://docs.astropy.org/en/stable/timeseries/lombscargle.html}, the LS periodogram provided a detailed analysis of quasi-periodic signals within the corresponding time series.
The WWZ is an improved version of the wavelet transform that is useful for detecting periodic or pseudo-periodic signals, even in uneven time series, and for assessing their persistence throughout the observation. The signal is simultaneously decomposed into both the frequency and time domains. This method uses wavelet functions to fit the observations rather than sinusoidal components. We perform a thorough WWZ analysis\footnote{https://github.com/skiehl/wwz} of the combined light curves over the entire observation period.

It is important to note that blazar light curves can exhibit both periodic and stochastic variability. The identification of quasi-periodic oscillations (QPOs) is attributed to systematic dynamics within the disc or jet. It is critical to make reliable estimates of the significance of any possible detection that could be misinterpreted as periodic peaks due to noise in the power spectrum. 
Therefore, we simulated 10000 light curves corresponding to the power spectral density of the blazar light curve using the same methodology\footnote{https://github.com/skiehl/lcsim} as described in \citet{2022ApJ...926L..35O}. By performing identical period analyses using WWZ and LS on the simulated light curves, as we did on the blazar light curve, we obtained significance levels of 95\% and 99\% for each frequency.

\begin{figure*}[bt!]
\centering
\includegraphics[width=0.95\textwidth]{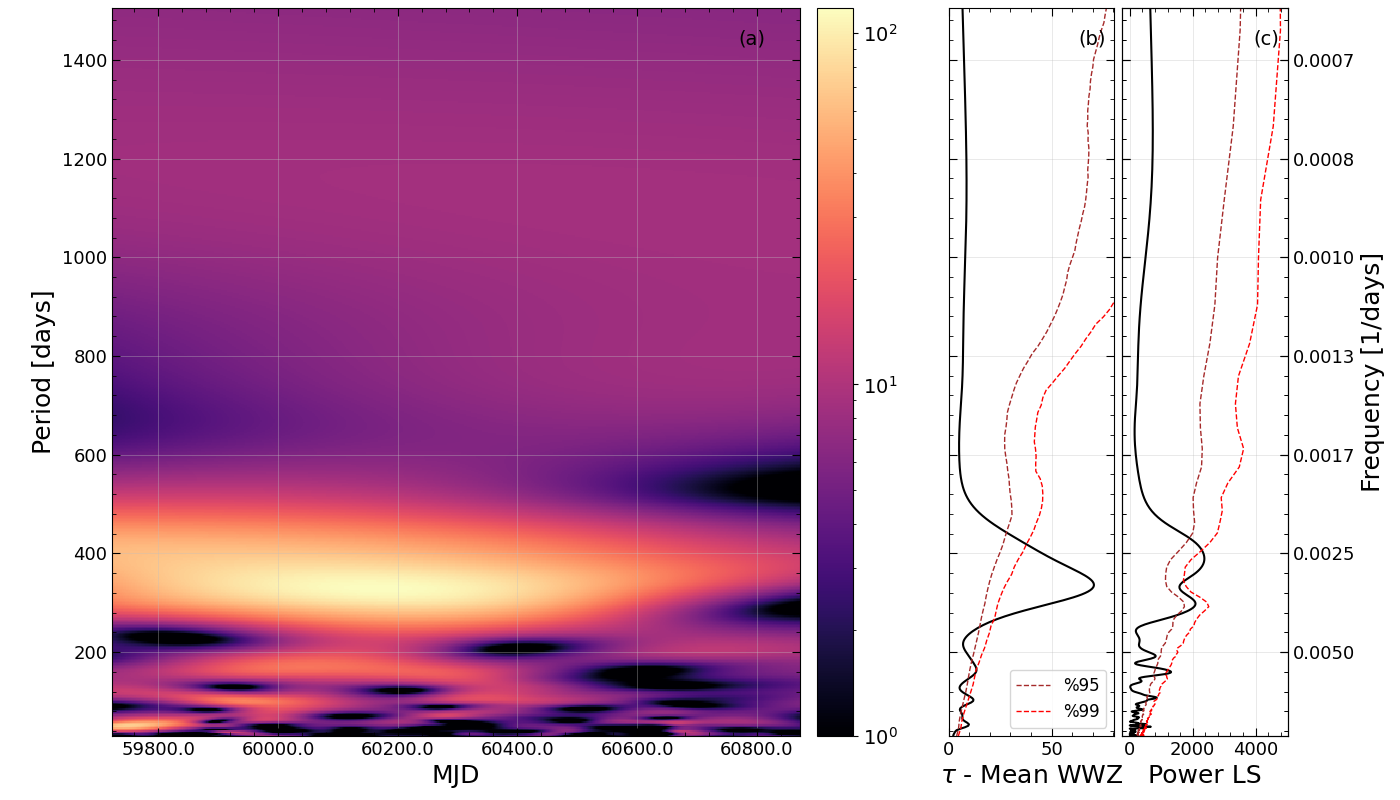}
\caption{(a) 2D plane contour plot of the WWZ power over time and frequency domain, (b) Time averaged WWZ power periodogram, and (c) the Lomb-Scargle periodogram of the combined $R$ band light curve. The dashed brown and red lines represent the 95\% and 99\% significance levels, respectively.}
\label{fig:wwz_ls}
\end{figure*}

The results of the WWZ and LS analyses for the combined $R$-band light curve are shown in Figure \ref{fig:wwz_ls}, including the WWZ power map corresponding to the periodicity and time, the periodogram of the time-averaged WWZ power, and the LS periodogram.
The WWZ analysis revealed a dominant signal at a periodicity of $337^{+133}_{-67}$ days in the WWZ periodogram, spanning the entire observation time in the WWZ power map and exceeding the significant level at 99\%. On the other hand, the LS periodogram exhibited two peak signals at $301\pm28$ days and $390\pm60$ days, with a combined signal at $390^{+55}_{-112}$ days. These signals exceeded the 95\% significance level, but did not reach the 99\% threshold.
Similar results were obtained for the combined V-band light curve as well as for the ZTF g- and r-band light curves alone. Our results suggest a possible recurrent behaviour in the optical emission of the blazar with a periodicity of about $340$ days.

\section{Discussion and Conclusions}
The most common explanation for the flux variations of the BL Lac type blazars during an outburst is the jet-in-shock scenario \citep[e.g.,][]{1985ApJ...298..114M, 1995ARA&A..33..163W, 2008Natur.452..966M}.
In this model, variability arises from a shock or blob propagating through relativistic plasma jets and interacting with the emitting regions. 
In addition, helical structures, precession or other geometric effects within the jets can change the Doppler factor of the emitting region resulting in flux variability \citep[e.g.,][]{1999A&A...347...30V, 2017Natur.552..374R}.
Hotspots or other emission regions on the accretion disc may contaminate the observed optical emission in the blazar during the low state due to the weakened contribution of the jets to the thermal emission \citep[e.g.,][]{1993ApJ...411..602C, 1993ApJ...406..420M, 1985ApJ...298..114M}. These mechanisms provide plausible explanations for the observed variations on timescales ranging from hours to months and years.

We present the results of our multi-band observations from 2020 to 2023 and ZTF observations from 2018 to 2023, studying the intraday, short-term, and long-term variability of the blazar 1E 1458.8+2249 in optical bands. 
Statistical tests on the seven intraday light curves indicate that the blazar has no significant variation on minute timescales. Combining our IDV results with those obtained in \citet{2023MNRAS.520.4118C}, the duty cycle (DC), which is defined as the fraction of nights showing variability out of the total number of nights \citep{1999A&AS..135..477R}, is calculated to be about 10\%. 
It's likely that HBLs, such as 1E 1458.8+2249, have lower DCs and smaller variability amplitudes than low- and intermediate-energy peaked blazars due to the prevention of the magnetic field \citep{1998A&A...329..853H, 1999A&AS..135..477R, 2011MNRAS.416..101G}. 
\citet{1995Ap&SS.234...49R} demonstrated that axial magnetic fields inhibit the formation of Kelvin-Helmholtz instabilities in jets spanning from sub-parsec to parsec scales. This prevention is effective if the magnetic fields exceed the critical value $B_c$, which is defined by the equation $B_c=[4\pi nm_ec^2(\Gamma^2-1)]^{1/2} \Gamma^{-1}$. Here, $n$ represents the local electron density, $m_e$ refers to the electron rest mass, and $\Gamma$ denotes the jet’s bulk Lorentz factor.
Higher magnetic fields ($B>B_c$) in HBLs would suppress the development of small-scale structures, hence decreasing the occurrence of microvariability in the optical light curves.
Nevertheless, the DC fraction is below the typical value of $\sim30-50$ observed in HBL objects \citep[e.g.,][]{2002A&A...390..431R, 2011MNRAS.416..101G}. 

On the other hand, the extreme variation of 0.97 magnitudes reported by \citet{2023MNRAS.520.4118C}, in an I-band IDV LC with an average magnitude of 16.43 is comparable to both the STV and even the LTV of the blazar. All other intraday light curves were taken in brighter states of blazar 1E 1458.8+2249, with an average magnitude between 15 and 16 mag. The lack of significant intraday variability detected in our study and in the literature suggests that the probability of detecting variability increasing with observation time alone may not sufficiently explain the absence of significant intraday variability. This makes it reasonable to consider alternative instability models for the IDV.

The blazar 1E 1458.8+2249 exhibits moderate STV and LTV, with an amplitude of up to 1.1 mag, and no flare is detected during the five years of observation. The multi-band LTV LCs show a strong correlation without any time lag for a 2-day time binning. This suggests that the optical emission region is likely to be co-spatial or that any spatial differences are too small to be detected with the optical datasets in this study. 
To understand the spectral variation of the source, we computed the spectral indices of the SEDs in the 62 nightly LCs ranging from 0.826 to 1.360 with an average of $1.128\pm0.063$ which are consistent with those obtained for HBLs within the range of 0.5 to 1.8 \citep[e.g.,][]{2012MNRAS.425.3002G, 2018ApJS..237...30M}. However, it is expected that the spectral index should be $\leq0.80$ based on simple pure synchrotron emission models and the spectral index of optically thin synchrotron-emitting plasma \citep{1995PASP..107..803U}. This suggests that variations in the contribution of different emission components to the optical emission of the blazar 1E 1458.8+224, such as thermal emission from the accretion disc or host galaxy, or non-thermal emission from different regions of the relativistic jets, are likely responsible for both the deviation in the spectral indices and the larger mean spectral index.
Analysis of the variation of the spectral indices with V-band magnitudes over the entire observation period shows a mild detectable BWB trend, consistent with the colour behaviour obtained from $VRI$-band data set for both IDV and LTV. However, ZTF spectral colours indicate stronger BWB trends.
The shock-in-jet model often cited in BL Lac studies can explain the mild BWB trend for both IDV and LTV in our study.

Our results suggest a possible recurrent behaviour in the optical emission of the blazar 1E 1458.8+2249 with a periodicity of about $340$ days. 
The jet itself is most likely driving the observed QPO in blazars, where it dominates the thermal emission and aligns closely with the observer's line of sight \citep[<\degree{10}][]{1995PASP..107..803U}.
It has been shown that if the blazar is part of a binary SMBH system it could also induce the jet precession effect caused by either closely orbiting binary black holes or warped accretion discs
\citep[e.g.,][]{2008Natur.452..851V, 2015MNRAS.453.1562G, 2016ApJ...832...47B}.
These mechanisms are mostly expected to generate QPOs  at year-like timescales. If the angle between the jet axis and the observer's line of sight is too small and the jet has a high Lorentz factor, the jet precession effect could lead to shorter periodicities, such as the observed QPO of 340 days
\citep[e.g.,][]{2016AJ....151...54S, 2016ApJ...832...47B, 2017ApJ...835..260Z}.
It is very likely that helical structures are common in blazars \citep{1999A&A...347...30V, 2004ApJ...615L...5R}. The Doppler boosting effect due to helical or non-ballistic motions of relativistic blobs or shocks within blazar jets could be responsible for the QPO in blazar LCs. 
\citep[e.g.,][]{1992A&A...255...59C, 2015ApJ...805...91M}. For the simplest leptonic one-zone model, due to the postulated helical motion of the blob, the viewing angle of the blob with respect to the line of sight ($\phi_{obs}$) changes periodically with time, and the Doppler factor ($\delta$) varies with viewing angle as 
$\delta=1/[\Gamma(1-\beta\cos{\theta(t)})]$ \citep{2017MNRAS.465..161S, 2018NatCo...9.4599Z}. Here,  
$\Gamma=1/\sqrt{1-\beta^2}$ is the bulk Lorentz factor of the blob motion with $\beta=\nu_{jet}/c$. Given this scenario, the periodicity in the rest frame of the blob is given by 
$P_{rf}={P_{obs}}/{(1-\beta\cos{\psi}\cos{\phi})}$. For typical values of 
the pitch angle of the helical path $\phi=$\degree{2}, 
the angle of the jet axis with respect to the line of sight $\psi=$\degree{2}
, and $\Gamma=4.6$ \citep{2007A&A...466...63W}, the rest frame periodicity is calculated as $\sim37$ yr for an observed periodicity of $P_{obs}=\sim340$ day. During one period, the blob traverses a distance of $D = c\beta P_{rf}\cos{\phi} \simeq 11$ pc.
In order to detect a statistically significant QPO in the present case, at least 4-5 cycles are required, thus  the blob would traverse $\sim55$ pc throughout the observation time. Under these assumptions, this length scale indicates a very highly curved jet structure to drive the QPO. In addition, a slow increasing trend in the first 5 cycles and then a continuous power trend close to zero can be seen in Figure \ref{fig:wwz_ls}, but a faster attenuation in the QPO amplitudes after peak is expected \citep{2022MNRAS.510.3641R}. Thus, this scenario is less likely for the QPO obtained from the dataset we have.
The periodicity driving mechanisms associated with the orbital motions of hot spots, spiral shocks, or other non-axisymmetric phenomena around the innermost stable circular orbit.  \citep[e.g.,][]{1993ApJ...406..420M, 2012MNRAS.423.3083M, 2013RAA....13..705H, 2019MNRAS.484.5785G} are most likely to produce much shorter periodicities than we have found.

Further multi-wavelength observations on diverse timescales, complemented by theoretical modelling of the underlying mechanisms of blazar variability, are needed to improve our understanding of the complex variability of the blazar 1E 1458.8+2249.


\begin{acknowledgement}
This study was supported by Scientific and Technological Research Council of Turkey (TUBITAK) under the Grant Number 121F427. The authors thank to TUBITAK for their supports. We thank the team of TUBITAK National Observatory (TUG) for a partial support in using the T60 and T100 telescopes with project numbers: 19BT60-1505, 22AT60-1907, 19AT100-1486, and 23AT100-2006.

\textit{Software/Python packages}:  
ccdproc \citep{2017zndo...1069648C}, 
Astropy \citep{2013A&A...558A..33A}, 
Numpy \citep{2020Natur.585..357H}, 
Matplotlib \citet{2007CSE.....9...90H}, 
Photutils \citep{2020zndo...4049061B},
WWZ \citep{2023ascl.soft10003K},
lcsim \citep{2023ascl.soft10002K}
\end{acknowledgement}



\paragraph{Data Availability Statement}
The data underlying this article are available in the article and in its online supplementary material.

\printendnotes

\bibliography{example}


\end{document}